LA-UR-

Title:

Author(s):

Intended for:

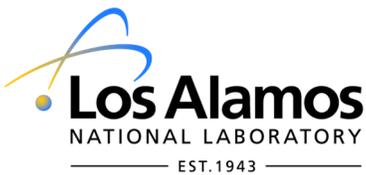

Form 836 (7/06)

# High Energy Activation Data Library (HEAD-2009)

Yu.A. Korovin[1], A.A. Natalenko[1], A.Yu. Konobeyev[1,2], A.Yu. Stankovskiy[1,3],

S.G. Mashnik[4]

[1]Obninsk state technical university for nuclear power engineering (INPE), Obninsk, Russia

[2]Association FZK-Euratom, Forschungszentrum Karlsruhe, Germany

[3]Advanced Nuclear Systems Institute, SCK·CEN, Belgium

[4]Los Alamos National Laboratory, Los Alamos, New Mexico 87545, USA

**Abstract**

A proton activation data library for 682 nuclides from $^1$H to $^{210}$Po in the energy range from 150 MeV up to 1 GeV was developed. To calculate proton activation data, the MCNPX 2.6.0 and CASCADE/INPE codes were chosen. Different intranuclear cascade, preequilibrium, and equilibrium nuclear reaction models and their combinations were used. The optimum calculation models have been chosen on the basis of statistical correlations for calculated and experimental proton data taken from the EXFOR library of experimental nuclear data. All the data are written in ENDF-6 format. The library is called HEPAD-2008 (High-Energy Proton Activation Data). A revision of IEAF-2005 neutron activation data library has been performed: A set of nuclides for which the cross-section data can be (and were) updated using more modern and improved models is specified, and the corresponding calculations have been made in the present work. The new version of the library is called IEAF-2009. The HEPAD-2008 and IEAF-2009 are merged to the final HEAD-2009 library.

# 1. Introduction

The sustainable development of the nuclear industry in many respects depends on solving the problem of nuclear waste management and disposal of long-lived fission products and minor actinides. As a strategy for handling radioactive waste, the transmutation process is offered. For transmutation, the use of dedicated installations is required. The accelerator driven transmutation technologies have been under development over the last 20 years. The accelerator-driven system (ADS) is composed of a blanket assembly of nuclear fuel containing reprocessed nuclear waste, a neutron spallation target irradiated by a high-energy proton beam and a proton accelerator.

For the calculation of the integral characteristics of the ADS blanket, such as the effective neutron multiplication factor and the reactivity coefficients, nuclear data in the energy range up to 20 MeV appear to be sufficient in most cases. But for modeling of particle transport, accumulation of spallation products, heating, radiating damage of the structural materials and many other similar problems in which it is necessary to consider separate elements of the system, relevant evaluated nuclear data in a wider energy range are necessary, typically from several keV up to several GeV. Thus, the new HEAD-2009 library can be useful in different applications in the field of fundamental and applied researches of the accelerator-driven systems and accelerator-driven transmutation technologies.

# 2. Activation data and structure of the HEAD-2009 library

The collected experimental activation data are contained in the EXFOR library [1]. A typical experiment for material activation is an expensive and time consuming task. For this reason a wide application of nuclear data based computer codes for simulation of high-energy nuclear reactions is required. Several data libraries and files containing activation cross-sections for a wide range of the nuclides and reactions were developed by different authors. The largest available proton and neutron activation data libraries are summarized in Tables 1 and Table 2, respectively. We can see a serious lack of activation data in the energy range above 200 MeV before our IEAF-2005 and HEPAD-2008 libraries were produced.

The current HEAD-2009 library consists of proton and neutron sub-libraries. Each sub-library includes a set of separate files for 682 nuclides. The files contain the cross-section data of the proton and neutron reactions with target nuclides in the energy range from 150 MeV to 1 GeV. The activation data are presented with a 5 MeV step in the energy range from 150 MeV to 250 MeV, and with 25 and 50 MeV steps in the energy ranges from 250 MeV to 600 MeV and from 600 MeV to 1 GeV, respectively. The resulting cross section data are processed in the ENDF-6 format making use of the MT=5 option, with the excitation functions stored in MF=3 and the product nuclide vectors in MF=6 (see Ref. [2] for details). Threshold reactions are written in MF=6, MT=5 starting from the first energy point equal to the reaction threshold.

## 3. Nuclear reaction models used to develop our HEPAD-2008 library

Generally, several codes implementing the quantum molecular dynamic (QMD) and many different semi-classical models are used to simulate nuclear reactions in the energy range 150 MeV – 1 GeV. For semi-classical models, the application field is defined by the energy of the particle or nuclei involved in the interactions. For the description of high-energy interactions (above 150 MeV), different versions of the intranuclear cascade (INC) model have a wide application. In the energy range from several dozen of MeV up to 150 - 250 MeV, various preequilibrium exciton models are used. At lower energies, the phenomenological evaporation model of Weisskopf [3] and the statistical model of Hauser and Feshbach [4] are usually applied. For heavy nuclei, along with evaporation, the fission process might take place. A general scheme of calculation high-energy nuclear reactions as implemented in the majority of modern codes is presented in Figure 1. An important feature influencing the results of simulations, apart from the parameters of each model implemented in a complex code, is the transition criteria between models describing different stages of reactions.

Production of light fragments with a mass number smaller than a predetermined value, $A_1$, from various nuclear reactions is described usually by different fragmentation models. Similarly, production of heavier fragments with mass numbers higher than another predetermined value, $A_2$, is described usually with fission models. A further de-excitation of fission fragments is described in the framework of different evaporation models. For the calculation of the proton activation data files, the MCNPX 2.6.0 [11] and CASCADE/INPE [12] codes have been used. For the description of nucleon-nuclide interactions, there are several models included in the general purpose radiation transport code MCNPX 2.6.0: There is a possibility to use different combinations of models. A list of physical models implemented in the MCNPX 2.6.0 and CASCADE/INPE codes and used in our HEPAD-2008 calculations is presented in Table 3.

The INCL4 and JINR/Dubna-based intranuclear cascade models have some recent improvements in comparison with older models: The number of quantum effects taken into account by these models is higher and the results obtained using these codes are often in a better agreement with the experimental data. These models and the ISABEL INC model are still under development nowadays, while the development of the Bertini model used in MCNPX was stopped in 1970-1980. The possibility for cluster formation during the cascade stage is considered by INCL4 and is accounted via coalescence of nucleons emitted during the INC into complex particles up to $^4$He by CEM03.01; this is a powerful feature of these models. The modified exciton model (MEM) of preequilibrium reactions developed at JINR, Dubna [20] allows transitions that reduce the number of excitons, in contrast to the Los Alamos multistep preequilibrium model (MPM) that uses only the "never come back" approximation [21]. For the calculation of the nuclear reaction cross-sections on light nuclides, only the MCNPX code was used. Moreover, the majority of the intranuclear cascade-preequilibrium-equilibrium models do not allow making reliable simulations for light nuclides with mass number A < 5. The only model allowing calculations for light nuclides (A < 5) within the MCNPX 2.6.0 is the Bertini INC; it was used in our calculations of the proton activation files for targets with A < 5. Note that the ancient version of the Fermi Breakup model used by CEM03.01 is has some problems and requires further improvements, a problem solved in the latest versions of CEM (see details in Ref. [28]).

### 4. Analysis of the EXFOR experimental data library

For cross-section calculations and the production of evaluated files, it is not enough to be guided only by analytical and expert estimations. Therefore, a detailed statistical analysis of correlations between the EXFOR experimental data and the results by MCNPX using eight different models listed in Table 3 (Bertini/Dresner, Bertini/ABLA, ISABEL/Dresner, ISABEL/ABLA, INCL4/Dresner, INCL4/ABLA, CEM03.01, and CASCADE) has been carried out.

All the suitable experimental data from the EXFOR library were chosen in our analysis. In total, about 4000 proton experimental points (measurements) for more than 1000 reactions with nuclides ranging from Z=6 to Z=84 and for incident particle energies from 150 to1000 MeV have been analyzed. The independent yield cross-sections for the following reactions were considered: (p,n), (p,α), (p,t), (p,d), (p,$^3$He) and others, designated here as (p,x). Distributions of the number of experimental measurements used in our analysis depending on the proton incident energy and on the target mass number are presented in Figure 2. We see that the most informative is the range from 150 to 450 MeV and also separate energy-points, like 500 MeV, 600 MeV, 660 MeV, 800 MeV, and 1 GeV. For $^{27}$Al and $^{208}$Pb, the EXFOR library contains the highest amount of experimental measurements.

## 5. Choice of the optimum models to calculate the HEPAD-2008 library files

We have used several statistical methods to determine the most appropriate models to use for the production of our library: a method of least squares, correlations, deviation factors, and regression analyses; but finally, the deviation factor analysis was chosen. For an estimation of correlations between the calculated and experimental data, the whole range of the experimental points (measurements) was divided into approximately equal sets based on the mass number of the target nuclides. Every set included about 400 experimental points, for which the (p,x) reaction cross-sections were calculated using the eight models mentioned above and then the calculated cross-sections were compared with the experimental value. This procedure allowed us to apply the frequency description of the models and to compensate partially the absence of experimental data for certain nuclides.

Within the deviation factor analysis, a product of the F- and H-factors was used as the estimated parameter:

$$F = 10^{\sqrt{\frac{1}{N}\sum_{i=1}^{N}\left(\lg\sigma_i^{\exp}-\lg\sigma_i^{calc}\right)^2}}$$

, where N is the total number of compared values, $\sigma_i^{\exp}$ are the experimental cross section values, and $\sigma_i^{calc}$ are the calculated cross section values;

$$H = \sqrt{\frac{1}{N}\sum_{i=1}^{N}\left(\frac{\sigma_i^{\exp}-\sigma_i^{cacl}}{\Delta\sigma_i^{\exp}}\right)^2}$$

, where $\Delta\sigma_i^{\exp}$ are the uncertainties of the experimental cross section values [29].

In this case, the F-factor allows estimating the correlation between calculated and experimental data, while the H-factor takes into account the experimental errors. Both these factors are widely used in the literature in such types of comparisons.

The following rejection criterion for the calculated and experimental data was used:
- the experimental point for which the ratio of the experimental toward the theoretical cross-sections of the residual nuclides yield drops out of the range 0.01÷100 was not taken into account;
- if at least one of the calculated cross-sections was equal to zero, the experimental point was excluded from the consideration;
- if the value of the error of the experimental cross-section was not been specified by the authors of the experiment, the experimental point was excluded from the consideration.

The results of our analysis are summarized in Table 4. Note that for the light nuclides from $^1$H to $^4$He, the deviation factor analysis was not performed, as the majority of models do not simulate reliably nuclear reactions on light nuclides (A<5).

For every set of mass numbers, an optimum model has been selected for further calculations. The selected models for all sets are presented in Table 5. These models have been used in our calculations of the HEPAD-2008 library files. The Bertini/Dresner model, which is the fastest among all INC models, has been chosen for the calculations in the target mass range A=56-59. It produces results very similar to the INCL4 model, but is about ten times faster.

The proton activation data files are produced in the ENDF-6 format from the results in the MCNPX 2.6.0 and CASCADE/INPE output files. The relative error, R, was calculated as the ratio of the estimated variance to the average value of the residual cross-section obtained running N Monte-Carlo histories [30]. It has been chosen as a rejection criterion which defined the possibility for inclusion of the proton cross-section data in the library files:

$$R = \frac{S_{\bar{x}}}{\bar{x}} = \left[\frac{\sum_{i=1}^{N} x_i^2}{\left(\sum_{i=1}^{N} x_i\right)^2} - \frac{1}{N}\right]^{1/2} = \left[\frac{1}{n} - \frac{1}{N}\right]^{1/2},$$

where i is the current number of the Monte-Carlo history, N is the total number of histories, $x_i$ is the average residual production cross-section, and n is the number of the Monte-Carlo histories which resulted in the production of the given residual.

The cross-section data were compiled to form the library files if the relative error did not exceed 20 %. Finally, a comparison of the HEPAD-2008 and EXFOR cross-section data has been made. In some cases, considerable divergences were found and eliminated partially, if the completeness of the experimental data allowed it. Several examples of a good agreement between the experimental data and our HEPAD-2008 evaluated files are presented in Figure 3.

**6. Updated version of the IEAF-2005 neutron activation nuclear data library**

The IAEF-2005 activation data library was prepared in 2005 as a part of the work on the high-energy activation data library development [31]. The library was developed using the MCNPX 2.5.d [32] and CASCADE/INPE [12] codes and contains neutron-induced activation and transmutation cross sections for target nuclides with Z=1 to Z=84 and for neutron energies up to 1 GeV.

The IEAF-2005 and HEPAD-2008 libraries were created using several codes and models, taking advantage of the improvements in the nuclear reaction models over the last several years. Improved models have been applied to calculate the proton activation data. The MCNPX 2.5.d and CASCADE/INPE codes were used for the IEAF-2005 development and the MCNPX 2.6.0 and CASCADE/INPE codes were used for the HEPAD-2008. Several major differences in the nuclear reaction models implemented in MCNPX 2.5.d and MCNPX 2.6.0 are summarized in Table 6.

As one can see from Table 6, the CEM model has been improved significantly. The CEM and INCL4 models are the most extensively developed and tested nowadays; therefore, they are expected to be among the most reliable. Let us note several basic changes in the models of CEM03.01 in comparison with its previous versions:

- new approximations for the total elementary cross sections have been developed;
- allows the possibility of cluster formation via coalescence of nucleons emitted during INC into complex particles up to $^4$He;
- the condition for transition from the INC stage of a reaction to preequilibrium was improved;
- the original Modified Exciton Model (MEM) of the multiparticle preequilibrium decay was significantly improved;
- the evaporation stage of reactions is calculated with an improved version of the Generalized Evaporation Model (GEM2) by Furihata [24];
- the fragmentation of excited light nuclei with A<13 produced after the INC is considered using the Fermi break-up model.

The models recommended on the basis of our statistical analysis for calculations of the neutron and proton activation data in the various mass ranges of target nuclei are summarized in Table 7. The statistical analysis was performed for the whole set of experimental proton data taken from the EXFOR library. We see that all changes in model recommendations concerning the choice of models in HEPAD-2008 in comparison with the older IEAF-2005 are connected with the changes and improvements in CEM03.01:

- as mentioned above, none of the models used currently in MCNPX except the Bertini INC provide the possibility of performing calculations for nuclei with A< 5; therefore the MCNPX interpolation table has been used;
- as it is seen from Tables 8, for the set of A=12-19, the ISABEL/Dresner model has been initially recommended for IAEF-2005, but a later estimation has shown that the CEM2k gives better values for the product of deviation factors in comparison with the ISABEL/Dresner choice;
- for the set of nuclides with A=125-181, the CEM2k and CASCADE models provided practically equal values for the product of deviation factors (the CEM2k model was finally chosen for the calculations because CASCADE does not predict production of some residual nuclei). From Table 9 we can see that CEM03.01 and CASCADE provide practically equal products of deviation factors; thus, we can conclude that CEM2K, CEM03.01, and CASCADE can be recommended with a similar degree of accuracy for this set of target masses;

- for the heavier nuclei, the CEM03.01 code is recommended now instead of the CASCADE code. In tables 8 and 9, the range of mass numbers A=182-209 is divided into two sub-sets with A=182-197 and A=206-209, respectively. When producing IEAF-2005, they were united into a single set, since in both sets the CASCADE model provided better values for the product of deviation factors in comparison with CEM2k. Now, for the set with A=182-197, the CASCADE and CEM03.01 codes provide approximately similar results, but for the set with A=206-209, the CEM03.01 code appears to work much better.

In accordance with the discussion above, the files of the IEAF-2005 neutron activation library have been updated using the CEM03.01 model for the nuclides with mass numbers A=5-22 (the Bertini INC is always used by MCNPX to calculate interactions with A<5 nuclei) and for A=206-209; for other target mass numbers, the IEAF-2005 library remains unchanged. On the whole, 23 files for light nuclei and 16 files for heavy nuclei have been updated in the present work.

In conclusion, to convince ourselves of the reliability in the chosen models, the cross-section values of (p,x) reactions for some isotopes of lead and bismuth have been compared with the experimental data. As we practically do not have at present reliable experimental activation cross sections for neutron-induced reactions at energies above 150 MeV, we compared our calculations for neutron-induced reactions with the corresponding available proton-induced measurements from EXFOR. The values of the dispersion between the proton experimental data from the EXFOR library and the corresponding neutron-induced calculations by CASCADE/INPE from IEAF-2005 and the corresponding neutron-induced results recalculated here by CEM03.01 using formula (1) are presented in Table 10.

$$S = \sqrt{\frac{\sum_{i=1}^{n}(\sigma_i^{exp} - \sigma_i^{calc})^2}{\sum_{i=1}^{n}(\sigma_i^{exp})^2}}, \tag{1}$$

where $\sigma_i^{exp}$ and $\sigma_i^{calc}$ are the experimental and calculated cross-section values and n is the number of experimental points for considered reactions.

As seen from Table 10, the results obtained with CEM03.01 model are in a better agreement with the experimental data almost for all these reactions. The CEM03.01 code to a lesser degree and CASCADE/INPE to a higher degree underestimate the residual cross-sections in comparison with the experimental data for all these nuclides and almost for all considered energies of projectiles.

## 7. Conclusions

The HEPAD-2008 proton activation data library was developed. The majority of the library files were calculated using the modified JINR/Dubna-based intranuclear cascade-exciton or cascade-evaporation models from the CEM03.01 or CASCADE/INPE codes. The remaining files were calculated using the MCNPX 2.6.0 radiation transport code: The INCL4 intranuclear cascade model was used in combination with the Dresner evaporation and RAL fission models. This choice of models has been dictated by the results of the comparison of the EXFOR experimental data with our model calculations. At the final stage, the HEPAD-2008 files have been corrected to the experimental data.

A revision of the IEAF-2005 neutron activation data library has been performed, a set of nuclides for which the cross-section data can be updated using available improved models were specified, and the corresponding calculations have been performed in this work. The files of the IEAF-2005 neutron activation library have been updated using the CEM03.01 model for the target mass numbers A=5-22 and A=206-209; the Bertini INC was used in MCNPX to calculate interactions with A<5 nuclei. In total, 23 files for light nuclei and 16 files for heavy nuclei have been updated. The new version of the library is called IEAF-2009.

The HEPAD-2008 proton activation nuclear data library and the updated IEAF-2009 neutron activation nuclear data library are merged into the HEAD-2009 High Energy Activation Data Library. The HEAD-2009 Library files can be obtained upon request from the authors.

## Acknowledgements

This work has been supported in part by the U.S. Department of Energy at Los Alamos National Laboratory under Contract DE-AC52-06NA25396.

We thank Dr. Roger L. Martz for a careful reading of our manuscript and useful suggestions.

**References**

1. O. Schwerer (editor) on behalf of the Nuclear Reaction Data Centres Network: EXFOR Formats Description for Users (EXFOR Basis), IAEA-NDS-206, June 2008.
2. M. Herman, ENDF-102 Data Formats and Procedures for the Evaluated Nuclear Data File ENDF-6, BNL-NCS-44945-05, 2005.
3. V. Weisskopf, Phys. Rev. 52 (1937) 295.
4. W. Hauser, H. Feshbach, Phys. Rev. 87 (1952) 366.
5. The European Activation File: EAF-2007, http://www.nea.fr/dbforms/data/eva/evatapes/eaf_2007/
6. A.Yu. Konobeyev, C.H.M. Broeders, U. Fischer, L. Mercatali, I. Schmuck, S.P. Simakov, The Proton Activation Data File PADF-2007, DOI: 10.1051/ndata:07352, in Proceedings of the International Conference on Nuclear Data for Science and Technology, April 22-27, 2007, Nice, France, editors O. Bersillon, F. Gunsing, E. Bauge, R. Jacqmin, and S. Leray, EDP Sciences, 2008, pp. 709-712.
7. JENDL-3.3, http://www.nea.fr/dbforms/data/eva/evatapes/jendl_33/
8. JEFF-3.1, http://www.nea.fr/html/dbdata/JEFF/
9. TENDL-2009, http://www.talys.eu/tendl-2009/
10. A.Yu. Konobeyev, Yu.A. Korovin, V.P. Lunev, V.S. Masterov, et al: Voprosy Atomnoi Nauki i Techniki (Problems of Nuclear Science and Technology), Series: Nuclear Data, 3-4 (1992) 55.
11. D.B. Pelowitz (editor), MCNPX User's Manual, Version 2.6.0, LA-CP-07-1473 (April 2008).
12. V.S. Barashenkov, Yu.A. Korovin, A.Yu. Konobeyev, V.N. Sosnin, Atomnaya Energiya, 87 (1999) 283.
13. S.G. Mashnik, K.K. Gudima, A.J. Sierk, M.I. Baznat, N.V. Mokhov, CEM03.01 User Manual, Los Alamos National Laboratory Report LA-UR-05-7321, Los Alamos, 2005.
14. H.W. Bertini, Low-Energy Intranuclear Cascade Calculation, Phys. Rev. 131 (1963) 1801.
15. H.W. Bertini, Intranuclear-Cascade Calculation of the Secondary Nucleon Spectra from Nucleon-Nucleus Interactions in the Energy Range 340 to 2900 MeV and Comparison with Experiments, Phys. Rev. 188 (1969) 1711.
16. Y. Yariv, Z. Fraenkel, Intranuclear Cascade Calculation of High-Energies Heavy-Ion Interactions, Phys. Rev. C20 (1979) 2227.
17. Y. Yariv, Z. Fraenkel, Intranuclear Cascade Calculation of High-Energies Heavy-Ion Collisions: Effects of Interactions between Cascade Particles, Phys. Rev. C24 (1981) 488.

18. A. Boudard, J. Cugnon, S. Leray, C. Volant, Intranuclear Cascade Model for a Comprehensive Description of Spallation Reaction data. Phys. Rev. C66 (2002) 044615.
19. V.S. Barashenkov, V.D. Toneev, Interactions of high energy particles and atomic nuclei with nuclei, Atomizdat, 1972 (in Russian).
20. K.K. Gudima, G.A. Ososkov, V.D. Toneev, Model for Pre-Equilibrium Decay of Excited Nuclei. Sov. J. Nucl. Phys. 21 (1975) 138; S.G. Mashnik, V.D. Toneev, MODEX - the Program for Calculation of the Energy Spectra of Particles Emitted in the Reactions of Pre-Equilibrium and Equilibrium Statistical Decays, Communication JINR P4-8417, Dubna, 1974 (in Russian).
21. R.E. Prael, M. Bozoian, Adaptation of the Multistage Preequilibrium Model for the Monte Carlo Method (I), Los Alamos National Laboratory Report LA-UR-88-3238, Los Alamos, 1988.
22. L. Dresner, EVAP, A Fortran Program for Calculation the Evaporation of Various Particles from Excited Compound Nuclei, Oak Ridge National Laboratory Report ORNL-TM-196, Oak Ridge, 1962.
23. A.R. Junghans, M. de Jong, H.-G. Clerc, A.V. Ignatyuk, G.A. Kudyaev, K.-H. Schmidt, Projectile-fragment yields as a probe for the collective enhancement in the nuclear level density, Nucl. Phys. A. 629 (1998) 635.
24. S. Furihata, Development of a generalized evaporation model and study of residual nuclei production, Ph.D. thesis, Tohoku University, Japan, 2003.
25. F. Atchison, Spallation and Fission in Heavy Metal Nuclei under Medium Energy Proton Bombardment in Targets for Neutron Beam Spallation Sources, Jul-Conf-34, Kernforschungsanlage Julich GmbH, 1980.
26. D.J. Brenner, R.E. Prael, J.F. Dicello, M. Zaider, Improved Calculations of Energy Deposition from Fast Neutrons, in Proceedings of the Fourth Symposium on Neutron Dosimetry, EUR-7448, Munich-Neuherberg, 1981; Los Alamos National Laboratory Report LA-UR-81-1694, Los Alamos, 1981.
27. N. Amelin, Physics and Algorithms of the Hadronic Monte-Carlo Event Generators. Notes for a Developer, CERN/IT/ASD Report CERN/IT/99/6, Geneva, Switzerland and JINR/LHE, Dubna, Russia; Geant4 User's Documents, Physics Reference Manual, 1998.
28. S.G. Mashnik, K.K. Gudima, R.E. Prael, A.J. Sierk, M.I. Baznat, N.V. Mokhov, CEM03.03 and LAQGSM03.03 Event Generators for the MCNP6, MCNPX, and MARS15 Transport Codes, Los Alamos National Laboratory Report LA-UR-08-2931, Los Alamos, 2008; E-print: arXiv:0805.0751v2; IAEA Report INDC(NDS)-0530, Distr. SC, Vienna, Austria, August 2008, p. 51.
29. R. Michel, P. Nagel, International Codes and Model Intercomparison for Intermediate Energy Activation Yields, NSC/DOC (97)-1, NEA/P&T No 14, NEA OECD, Paris, 1997; http://www.nea.fr/html/science/docs/1997/nsc-doc97-1/.
30. X-5 Monte Carlo Team, MCNP — A General Monte Carlo N-Particle Transport Code, Volume I: Overview and Theory, Los Alamos National Laboratory Report LA-UR-03-1987, Los Alamos, 2003.
31. Yu. Korovin U. Fischer, A. Konobeyev, A. Natalenko, G. Pilnov, A. Stankovskiy, A. Tikhonenko, Evaluation of activation nuclear data in the energy region 150MeV to 1 GeV, DOI: 10.1051/ndata:07236, in Proceedings of the International Conference on Nuclear Data for Science and Technology, April 22-27, 2007, Nice, France, editors O.

Bersillon, F. Gunsing, E. Bauge, R. Jacqmin, and S. Leray, EDP Sciences, 2008, pp. 1175-1178.
32. J.S. Hendricks, G.W. McKinney, L.S. Waters, T.L. Roberts, H.W. Egdorf, J.P. Finch, H.R. Trellue, E.J. Pitcher, D.R. Mayo, M.T. Swinhoe, S.J. Tobin, J.W. Durkee, F.X. Gallmeier, J.-C. David, MCNPX extensions version 2.5.0, Los Alamos National Laboratory Report LA-UR-05-2675, Los Alamos, 2005.

**Table 1**

The largest available evaluated proton activation nuclear data libraries.

| Library | Number of files / Nuclear charge number range | Primary proton energy range |
|---|---|---|
| The European Activation File, EAF-2007 [5] | 816 / 1-100 | up to 60 MeV |
| Proton Activation Data File, PADF-2007 [6] | 2355 / 12-88 | up to 150 MeV |
| JENDL High Energy File 2007, JENDL/HE-2007 [7] | 106 / 1-95 | up to 3 GeV |
| Joint Evaluated Fission and Fusion File, JEFF-3.1 [8] | 26 / 20-83 | up to 200 MeV |
| TALYS-based Evaluated Nuclear Data Library, TENDL-2009 [9] | 2375 / 6-110 | up to 200 MeV |
| **High-Energy Proton Activation Data, HEPAD-2008** | **682 / 1-84** | **up to 1 GeV** |

**Table 2**

The largest available evaluated neutron activation nuclear data libraries.

| Library | Number of files / Nuclear charge number range | Primary proton energy range |
|---|---|---|
| The European Activation File, EAF-2007 | 816 / 1-100 | up to 60 MeV |
| JENDL High Energy File 2007, JENDL/HE-2007 | 106 / 1-95 | up to 3 GeV |
| Joint Evaluated Fission and Fusion File, JEFF-3.1/A | 774 / 1-100 | up to 20 MeV |
| Medium Energy Nuclear Data Library, MENDL-2 [10] | 505 / 13-84 | up to 100 MeV |
| TALYS-based Evaluated Nuclear Data Library, TENDL-2009 | 2375 / 6-110 | up to 200 MeV |
| **The Intermediate Energy Activation File, IEAF-2005** | **682 / 1-84** | **up to 1 GeV** |

**Table 3**
Physical models of the MCNPX 2.6c and CASCADE/INPE codes used in our calculations.

| Model | Code | Nuclear reaction models | | | | |
|---|---|---|---|---|---|---|
| | | INC | Preequilibrium | Equilibrium | Fission | Fragmentation |
| Bertini/Dresner | MCNPX 26C | Bertini [14, 15] | MPM [21] | Dresner [22] | RAL [25] | Fermi Breakup Model [26] |
| Bertini/ABLA | MCNPX 26C | | | ABLA [23] | | |
| ISABEL/Dresner | MCNPX 26C | ISABEL [16,17] | | Dresner [22] | RAL [25] | |
| ISABEL/ABLA | MCNPX 26C | | | ABLA [23] | | |
| INCL4/Dresner | MCNPX 26C | INCL4 [18] | | Dresner [22] | RAL [25] | |
| INCL4/ABLA | MCNPX 26C | | | ABLA [23] | | |
| CEM03.01 [13] | MCNPX 26C | JINR/Dubna [19] (modified) | MEM [20] | GEM2 [24] | | Fermi Breakup (modified) [27] |
| CASCADE | CASCADE/ INPE | JINR/Dubna | - | JINR/Dubna (Weisskopf model) | JINR/Dubna | do not taken in to account |

**Table 4**
Values of the product of the F- and H-deviation factors for different sets of the target mass numbers (the best results are shown with underlined bold numerals).

| Sets of the target mass number | Bertini/ ABLA | Bertini/ Dresner | CEM03.01 | INCL4/ ABLA | INCL4/ Dresner | ISABEL/ ABLA | ISABEL/ Dresner | CASCADE |
|---|---|---|---|---|---|---|---|---|
| 12-19 | 0.024 | 0.022 | **0.008** | 0.033 | 0.030 | 0.016 | 0.015 | |
| 23-27 | 0.018 | 0.006 | 0.019 | 0.020 | **0.006** | 0.024 | 0.009 | 0.017 |
| 28-55 | 0.020 | 0.022 | 0.011 | 0.014 | 0.014 | 0.019 | 0.021 | **0.008** |
| 56-59 | 0.015 | **0.011** | 0.013 | 0.020 | **0.011** | 0.020 | 0.013 | 0.018 |
| 60-89 | 0.023 | 0.025 | 0.008 | 0.019 | 0.020 | 0.014 | 0.014 | **0.005** |
| 90-124 | 0.029 | 0.011 | 0.027 | 0.015 | **0.008** | 0.018 | 0.012 | 0.009 |
| 125-181 | 0.017 | 0.019 | 0.011 | 0.017 | 0.018 | 0.015 | 0.014 | **0.010** |
| 182-197 | 0.023 | 0.024 | **0.004** | 0.022 | 0.025 | 0.013 | 0.015 | **0.004** |
| 206-209 | 0.020 | 0.025 | **0.006** | 0.015 | 0.022 | 0.012 | 0.018 | 0.009 |

**Table 5**
The models applied to calculate the HEPAD-2008 library files.

| Nuclide set | Reference model |
|---|---|
| 1-H-1 – 2-He-4 | Bertini INC |
| 3-Li-6 – 10-Ne-22 | CEM03.01 |
| 11-Na-23 – 13-Al-27 | INCL4/Dresner |
| 12-Mg-28 – 27-Co-55 | CASCADE |
| 29-Cu-56 – 28-Ni-59 | Bertini/Dresner |
| 26-Fe-60 – 40-Zr-89 | CASCADE |
| 38-Sr-90 – 54-Xe-124 | INCL4/ Dresner |
| 50-Sn-125 – 75-Re-181 | CASCADE |
| 72-Hf-182 – 84-Po-210 | CEM03.01 |

**Table 6**
Some major differences in the nuclear reaction models implemented in the MCNPX 2.5.d and MCNPX 2.6.0 codes.

| Model title | Code | Nuclear reaction models | | | | |
|---|---|---|---|---|---|---|
| | | Intranuclear cascade | Preequilibrium | Evaporation | Fission | Fragmentation |
| CEM2k | MCNPX 2.5.d | JINR/Dubna (modified) | MEM (modified) | Evaporation, from MEM; fission, with the RAL model | | Not considered |
| CEM03.01 | MCNPX 2.6.0 | JINR/Dubna (modified) | MEM (modified) | GEM2 | | Fermi Breakup model (modified) |

**Table 7**
Models recommended on the basis of our statistical analysis and used in the IEAF-2005 and HEPAD-2008 libraries calculations.

| Target mass range | Models used for the IEAF-2005 calculations | Models used for the HEPAD-2008 calculations |
|---|---|---|
| 1-H-1 – 2-He-4 | MCNPX interpolation tables | Bertini INC |
| 3-Li-6 – 10-Ne-22 | ISABEL/ Dresner | CEM03.01 |
| 11-Na-23 – 13-Al-27 | INCL4/Dresner | INCL4/Dresner |
| 12-Mg-28 – 27-Co-55 | CASCADE | CASCADE |
| 29-Cu-56 – 28-Ni-59 | Bertini/ Dresner | Bertini/Dresner |
| 26-Fe-60 – 40-Zr-89 | CASCADE | CASCADE |
| 38-Sr-90 – 54-Xe-124 | INCL4/ Dresner | INCL4/ Dresner |
| 50-Sn-125 – 75-Re-181 | CEM2K | CASCADE |
| 72-Hf-182 – 84-Po-210 | CASCADE | CEM03.01 |

**Table 8**

Values of the product of deviation factors H and F for different sets of the target mass numbers for the IAEF-2005 library (the best results are shown with bold numerals).

| Target mass number sets | Bertini/ ABLA | Bertini/ Dresner | CEM2K | INCL4/ ABLA | INCL4/ Dresner | ISABEL/ ABLA | ISABEL/ Dresner | CASCADE |
|---|---|---|---|---|---|---|---|---|
| 12-19 | 0.022 | 0.020 | **0.014** | 0.030 | 0.028 | 0.016 | **0.014** | |
| 125-181 | 0.018 | 0.019 | **0.011** | 0.018 | 0.018 | 0.015 | 0.014 | **0.010** |
| 182-197 | 0.019 | 0.023 | 0.009 | 0.018 | 0.026 | 0.011 | 0.015 | **0.005** |
| 206-209 | 0.017 | 0.021 | 0.020 | 0.013 | 0.019 | 0.010 | 0.015 | **0.008** |

**Table 9**

Values of the product of deviation factors H and F for different sets of the target mass numbers for the HEPAD-2008 library (the best results are shown with bold numerals).

| Target mass number sets | Bertini/ ABLA | Bertini/ Dresner | CEM03.01 | INCL4/ ABLA | INCL4/ Dresner | ISABEL/ ABLA | ISABEL/ Dresner | CASCADE |
|---|---|---|---|---|---|---|---|---|
| 12-19 | 0.024 | 0.024 | **0.008** | 0.033 | 0.030 | 0.016 | 0.015 | |
| 125-181 | 0.017 | 0.019 | **0.011** | 0.017 | 0.018 | 0.015 | 0.014 | **0.010** |
| 182-197 | 0.023 | 0.024 | **0.004** | 0.022 | 0.025 | 0.013 | 0.015 | **0.004** |
| 206-209 | 0.020 | 0.025 | **0.006** | 0.015 | 0.022 | 0.012 | 0.018 | 0.009 |

**Table 10**
Dispersion values between the proton-induced experimental data from EXFOR and neutron-induced calculations by CASCADE/INPE from IEAF-2005 and the current CEM03.01 results for neutron-induced reactions, respectively.

| Nuclear reaction | IEAF-2005 (CASCADE/INPE) dispersion values | CEM03.01 model dispersion values | n |
|---|---|---|---|
| 82-PB-208(P,X)1-H-1 | 1.794 | *0.935* | 1 |
| 82-PB-208(P,X)1-H-2 | 0.957 | *0.066* | 1 |
| 82-PB-208(P,X)2-H-4 | 0.875 | *0.383* | 1 |
| 82-PB-208(P,X)2-HE-3 | 0.588 | *0.466* | 31 |
| 83-BI-209(P,X)2-HE-3 | *0.652* | 1.672 | 7 |
| 83-BI-209(P,X)2-HE-4 | 0.452 | *0.450* | 6 |
| 83-BI-209(P,X)29-CU-64 | *0.345* | 0.583 | 1 |
| 83-BI-209(P,X)33-AS-72 | 0.608 | *0.594* | 1 |
| 83-BI-209(P,X)33-AS-74 | 0.653 | *0.597* | 1 |
| 83-BI-209(P,X)33-AS-76 | 0.917 | *0.826* | 1 |
| 83-BI-209(P,X)33-AS-77 | 0.960 | *0.859* | 1 |
| 83-BI-209(P,X)35-BR-82 | 0.894 | *0.605* | 1 |
| 83-BI-209(P,X)35-BR-83 | 0.959 | *0.503* | 1 |
| 83-BI-209(P,X)37-RB-86 | 0.927 | *0.681* | 1 |
| 83-BI-209(P,X)38-SR-82 | - | *0.694* | 1 |
| 83-BI-209(P,X)38-SR-89 | 0.956 | *0.812* | 1 |
| 83-BI-209(P,X)38-SR-90 | 0.979 | *0.811* | 1 |
| 83-BI-209(P,X)42-MO-99 | 0.974 | *0.845* | 1 |
| 83-BI-209(P,X)46-PD-109 | 0.972 | *0.884* | 1 |
| 83-BI-209(P,X)47-AG-111 | 0.974 | *0.827* | 1 |
| 83-BI-209(P,X)47-AG-112 | 0.969 | *0.723* | 1 |
| 83-BI-209(P,X)47-AG-113 | - | *0.855* | 1 |
| 83-BI-209(P,X)48-CD-115 | 0.954 | *0.669* | 1 |

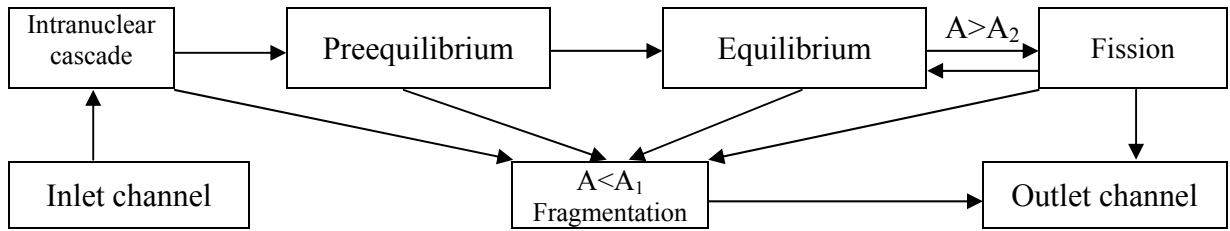

**Fig. 1.** A general scheme of the high-energy nuclear reaction calculation by our codes.

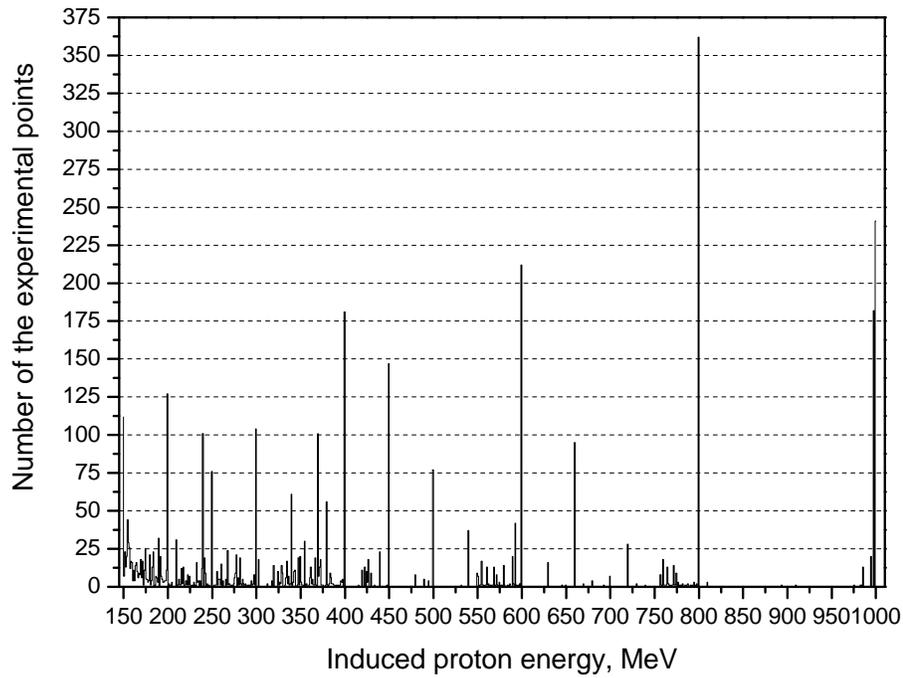

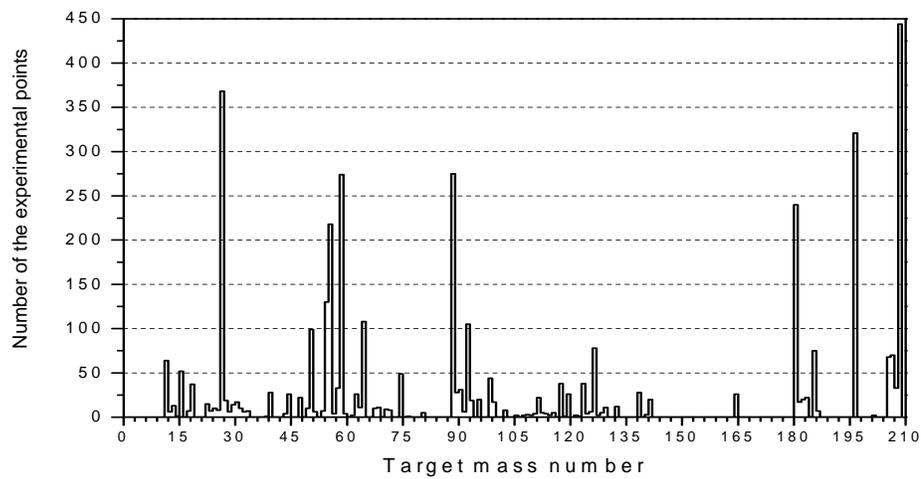

**Fig. 2.** Distributions of the available experimental measurements used in our analysis as functions of the proton bombarding energy (upper plot) and of the target mass number (lower plot).

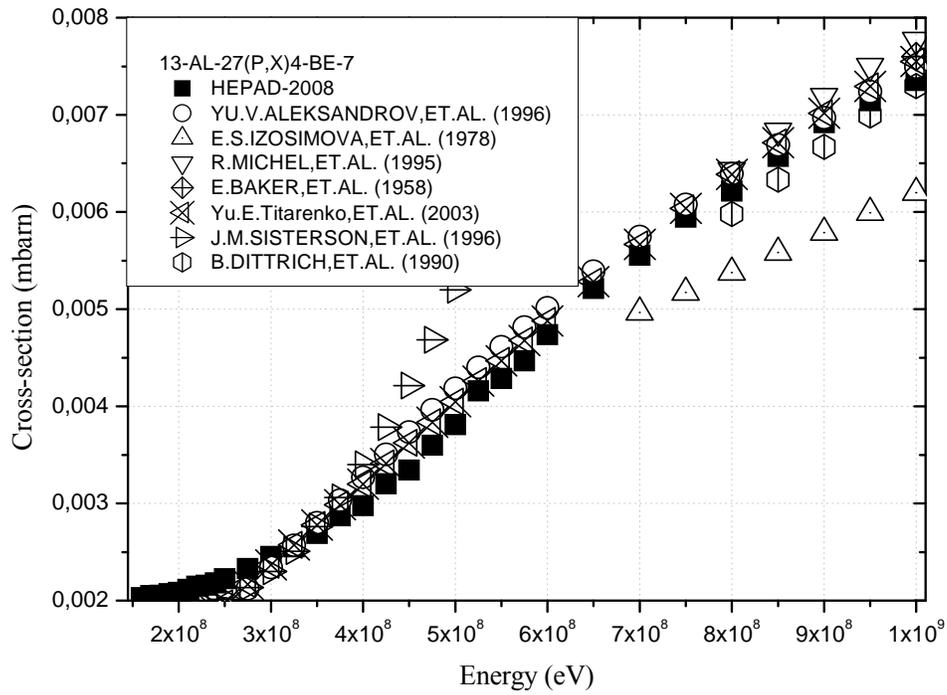

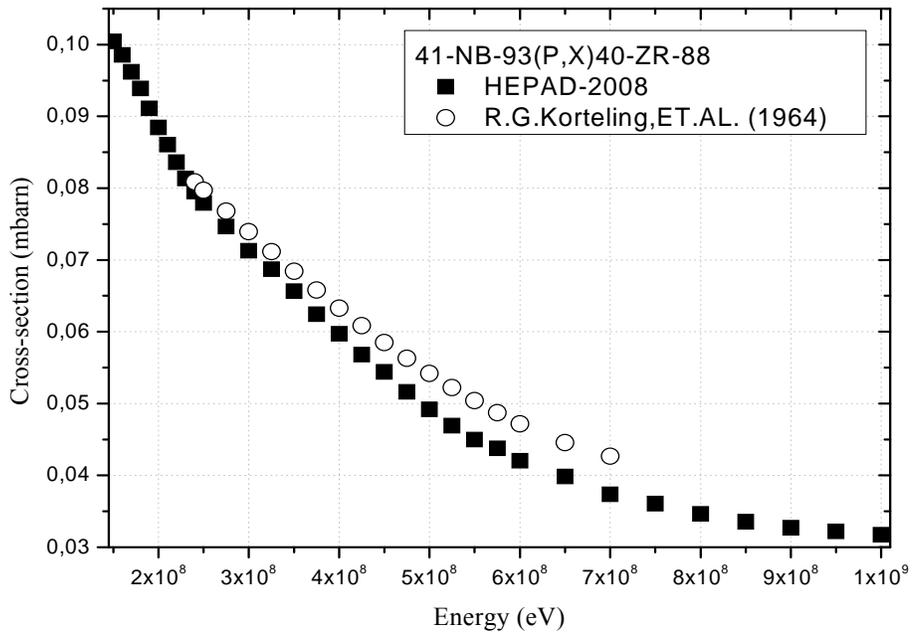

**Fig. 3.** Examples of excitation functions for several reactions comparing the available experimental data from EXFOR with our calculations from the HEPAD-2008 library.

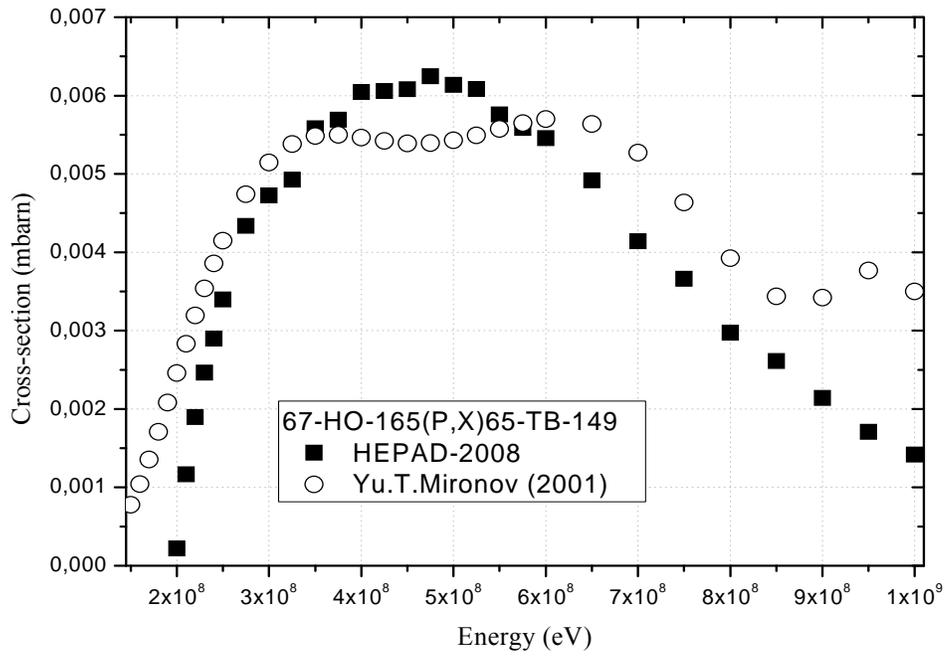

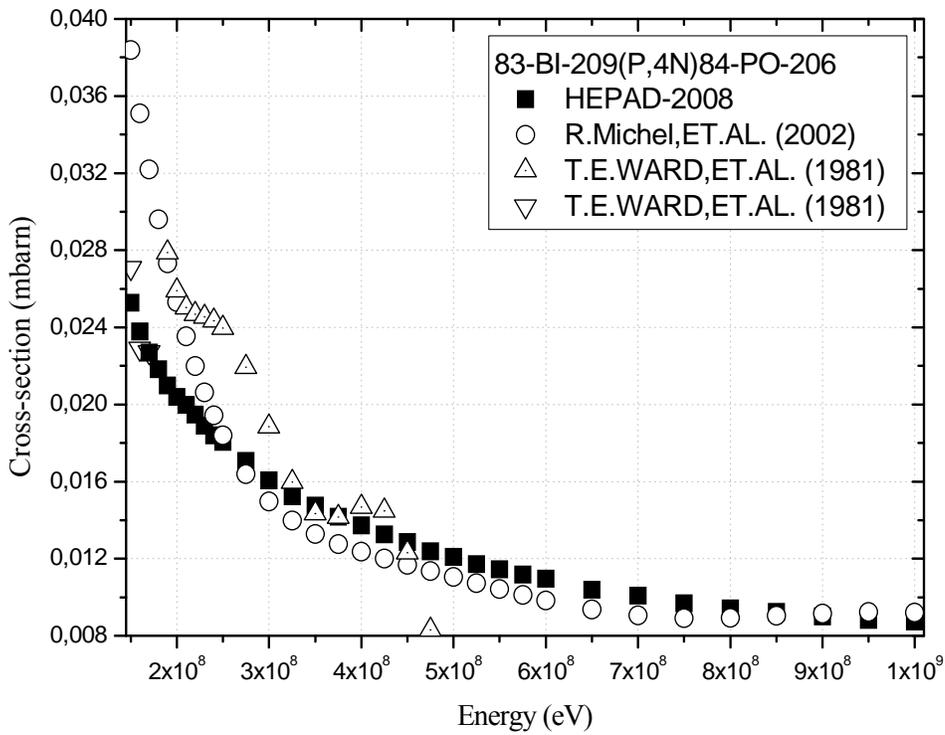

**Continuation of Fig. 3.**